# Comment on "Optical Response of Strongly Coupled Nanopraticles in Dimer Arrays" (Phys. Rev. B 71(4), 045404, 2005).

<'>

Vadim A. Markel

*Departments of Radiology and Bioengineering, University of Pennsylvania, Philadelphia, PA 19104*

(Dated: August 15, 2018)

I have re-calculated the extinction spectra of aggregates of two silver nanospheres shown in Figs. 2 and 3 of Ref. 8. I have used the approximate method of images according to Ref. 8 and an exact numerical technique. I have found that the three sets of data (those I have obtained by the method of images, the numerical results, and the results published in Ref. 8) do not coincide. In this Comment, I discuss the reasons for these discrepancies and the general applicability of the method of images to the quasi-static electromagnetic problem of two interacting nanospheres.

The method of images (MOI) is a powerful tool for solving electrostatic problems [1]. In particular, it has been used to calculate the electrostatic force between two charged spheres [2]. It was shown that the force can be well approximated by the Coulomb formula when the spheres are far apart. However, as the spheres approach each other, they can not be effectively replaced by point charges and the Coulomb formula overestimates the actual force. The MOI was recently extended beyond the electrostatics [3, 4, 5, 6, 7, 8]. However, this generalization was subject of some controversy [9, 10]. In particular, I have argued that the MOI, as used in the above references, is not a physically justifiable approximation and, therefore, can not be used for calculating electromagnetic responses of interacting spheres at finite frequencies [9]. I have further argued that the formulas derived in [3, 4, 5, 6, 7, 8] do not provide accurate results even within electrostatics, when the MOI is, in principle, applicable. The authors of Ref. [10] argued that the MOI is an accurate approximation at sufficiently low frequencies, e.g., for $f < 1$GHz (this claim will be discussed below). However, in Ref. [8], which is the subject of this Comment, the MOI is used for the Drude dielectric function in the spectral range from 250nm to 1500nm, i.e., at much larger electromagnetic frequencies. I have recalculated the data shown in Figs. 2,3 of Ref. [8] using the MOI as it is defined in [3, 4, 5, 6, 7, 8]. I have also calculated the relevant quantities using the exact method (e.g., see [11]). I have found that the three sets of data (i.e., the data shown in Figs. 2,3 of Ref. [8], the data I've obtained according the MOI using the same formulas as in Ref. [8], and the exact results) do not coincide. The deviations are significant. This and other points relevant to the applicability of the MOI are discussed in this Comment.

First, we specify the dielectric function used in calculations. The expression given in Ref. [8] is

$$\epsilon = \epsilon_h - (\epsilon_s - \epsilon_h)\omega_p^2/[\omega(\omega + i\gamma)] , \qquad (1)$$

where $\omega_p = 1.72 \times 10^{16}$rad/sec, $\epsilon_h = 5.45$, $\epsilon_s = 6.18$, and the relaxation constant $\gamma$ is size-dependent:

$$\gamma = v_F/\ell + 2v_F/d , \qquad (2)$$

where $v_F = 1.38 \times 10^6$m/sec is the Fermi velocity, $\ell = 52$nm is the electron free path and $d$ is the sphere diameter. All quantities give good approximation for silver. Since the electromagnetic frequency is expressed in the units of energy in Figs. 2,3 of Ref. [8], we re-write the expression (1) as

$$\epsilon = \epsilon_h + \frac{E_p^2}{E(E + i\Gamma)} , \qquad (3)$$

where $E = \hbar\omega$, $E_p = 9.68$eV and

$$\Gamma = 0.00181 \times E_p(1 + 2\ell/d) . \qquad (4)$$

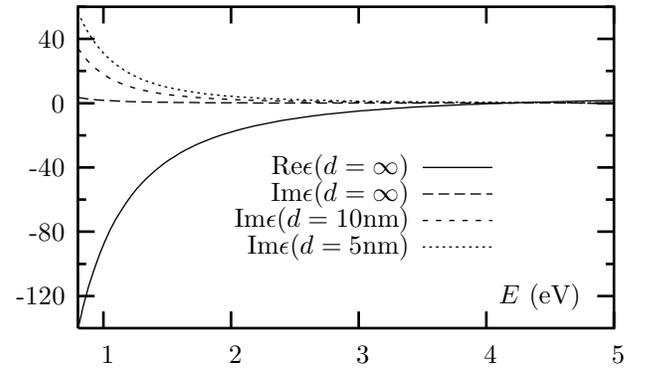

Fig. 1. Size-dependent dielectric function for different values of the sphere radius $d$ in the spectral range of interest.

The dielectric function defined by these equations is shown in Fig. 1 for different values of the sphere diameter $d$. It should be noted that an analogous graph is also shown in Fig. 1 of Ref. [8] for $d = 10$nm. While the real part of the dielectric function is qualitatively similar in both figures, the imaginary parts (for $d = 10$nm) are very different. By comparison with other curves in Fig. 1, I infer that the dashed curve shown in Fig. 1 of Ref. [8] corresponds to the case $d \to \infty$ rather than to $d = 10$nm, as claimed.

The mathematical formalism of MOI [3, 4, 5, 6, 7, 8] can be summarized as follows. The polarizability of each sphere in a two-sphere aggregate (radius of each sphere

is $d$ and the surface-to-surface separation is $\sigma$) is given by

$$\alpha = \frac{(d/2)^3}{3} \sum_{n=1}^{\infty} \frac{F_n}{s + s_n^{(L,T)}} \,, \qquad (5)$$

where $s = \epsilon_m/(\epsilon - \epsilon_m)$ is the spectral parameter of the theory, with $\epsilon_m$ being the dielectric function of the transparent matrix in which the spheres are embedded, the upper index $L$ corresponds to polarization of the external field parallel to the axis of symmetry of two spheres and the index $T$ corresponds to orthogonal polarization. The above formula is quite general and is not a subject of controversy. The values of $F_n$ and $s_n^{(L,T)}$ can be, in principle, found numerically. The potential advantage of the theoretical developments of Refs. [3, 4, 5, 6, 7, 8] is that it provides approximate analytical expressions for these quantities:

$$F_n = 4n(n+1)\sinh^3 a \exp[-(2n+1)a] \,, \qquad (6)$$
$$s_n^{(L)} = \frac{1}{3}\{1 - 2\exp[-(2n+1)a]\} \,, \qquad (7)$$
$$s_n^{(T)} = \frac{1}{3}\{1 + \exp[-(2n+1)a]\} \,, \qquad (8)$$
$$a = \ln\left[1 + \frac{\sigma}{d} + \sqrt{\frac{\sigma}{d}\left(2 + \frac{\sigma}{d}\right)}\right] \,. \qquad (9)$$

However, I have previously argued that (i) the values of these coefficients can not be, in principle, found from the MOI, even approximately, and (ii) the above expressions are inconsistent with the electrostatic limit [9]. Therefore, the formulas (6)-(8) do not provide a physically meaningful approximation. This is illustrated in Figs. 2 and 3 below.

We note that a slight change of notations has been adopted. Thus, the factors $F_n$ defined in (6) differ from those of Refs. [3, 4, 5, 6, 7, 8] by the overall factor of $-3$ and the spectral parameter $s = \epsilon_m/(\epsilon - \epsilon_m)$ by the factor of $-1$. This corresponds to the more conventional notations [12, 13, 14]. In particular, oscillator strengths (6) satisfy the sum rule $\sum_{n>0} F_n = 1$. Also, the expression for the polarizability given in Ref. [8] contains an extra factor of $\epsilon_m$ compared to Eq. (5). Although the inclusion of this factor in the expression for the polarizability is incorrect, as can be easily seen in the limit $\sigma/d \to \infty$, it does not change any of the spectral lineshapes.

In what follows, we consider only the results for the polarization of the external field being parallel to the axis of symmetry of two spheres, since the multipole interaction is strongest in this case. I have used the dielectric function defined above to calculate the extinction cross-section of the bisphere aggregate for the same sets of parameters as in Figs. 2,3 of Ref. [8]. Namely, the dielectric constant of the matrix was $\epsilon_m = (1.61)^2$, the sphere diameters were chosen to be $d = 5$nm (Fig. 2) and $d = 10$nm (Fig. 3), and the ratio $\sigma/d$ was 0.1 and 0.3 (Fig. 2) and 0.05, 0.15, 0.25 and 0.35 (Fig. 3). The extinction in Ref. [8] was plotted in arbitrary units and not defined precisely. Therefore, I plot the quantity $(E/3)\mathrm{Im}\sum_{n>0} f_n/(s+s_n)$, where $E$ is expressed in electron-volts. This quantity differs from the actual extinction cross section only by a constant factor, and I have found that it has approximately the same numerical values as the data shown in Figs. 2,3 of Ref. [8].

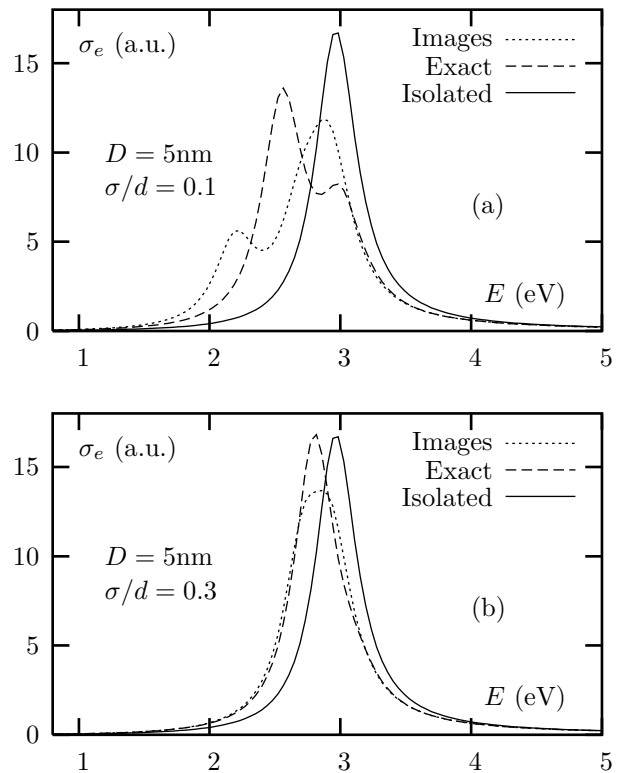

Fig. 2. Extinction spectra for a two-sphere aggregate in a transparent matrix with $\epsilon_m = 1.61^2$ and $d = 5$nm. Polarization of the external field is parallel to the axis of symmetry. The exact result are compared to MOI and to the spectra for isolated (non-interacting) spheres.

We first discuss Fig. 2 which is analogous to Fig. 2 of Ref. [8], except that data for orthogonal polarization are not shown. The exact spectra were calculated by the method of Ref. [11] with the maximum multipole order $L = 400$ and convergence was verified by doubling this number. The MOI calculations were carried out by truncating summation in (5) at $n = 400$, which is much more than is needed for convergence. First, it should be noted that the shape of the curves that I have obtained by the MOI are markedly different from those shown in Fig. 2 of Ref. [8]. In particular, the second, low-energy peak in the extinction spectrum for $\sigma/d = 0.1$ is significantly less pronounced in Ref. [8] than in my data. A noticeable deviation is also visible in the case $\sigma/d = 0.3$. A possible cause of this discrepancy is discussed below. More importantly, the MOI curves in both cases differ from the exact result. The difference is very apparent at the smaller separation ($\sigma/d = 0.1$), and still visible at the relatively large separation $\sigma/d = 0.3$.

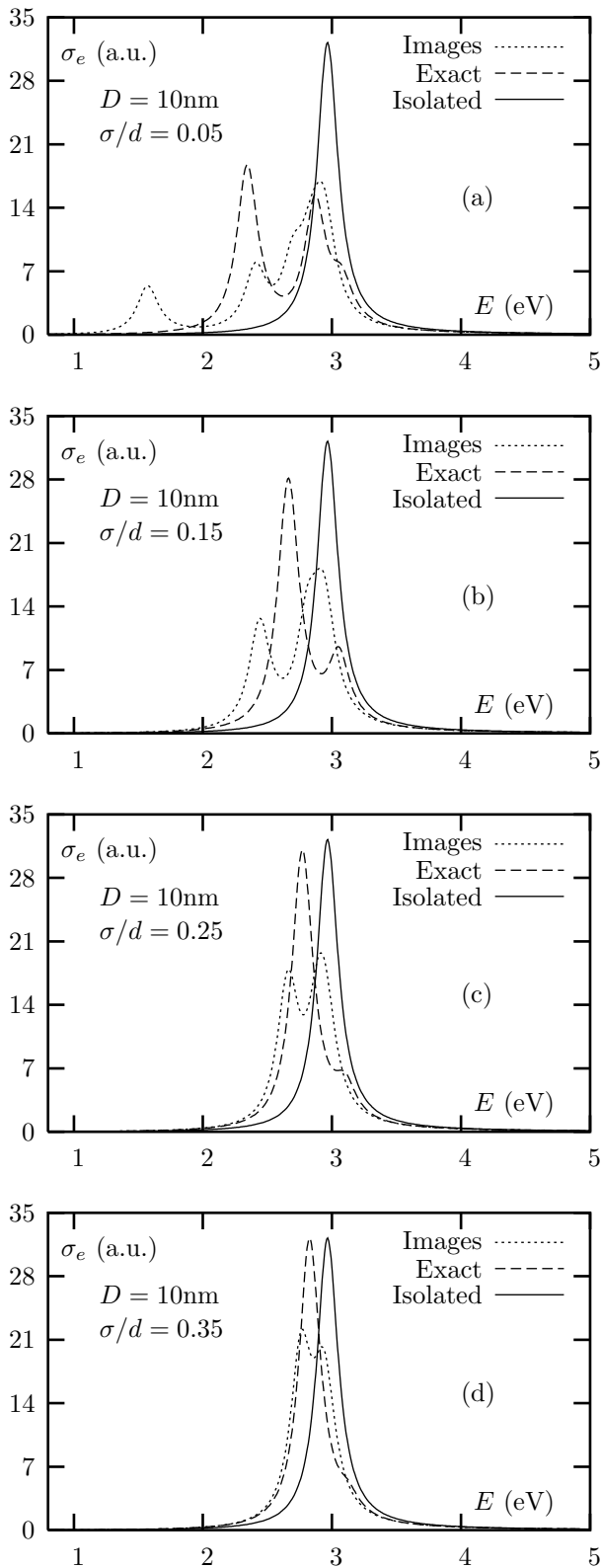

Fig. 3. Same as in Fig. 2, but for $d = 10$nm and a different selections of the ratio $\sigma/d$.

The spectra in Fig. 2 are characterized by very strong relaxation, because the ratio $2\ell/d$ is in this case of the order of 20. Thus, the finite size contribution to the relaxation constant is approximately 20 times larger than the respective constant in bulk. We then consider Fig. 3 ($d = 10$nm), where the relaxation is not as strong. The results are shown in Fig. 3 which corresponds to Fig. 3 of Ref. [8] with the exception that the results for orthogonal polarization are not shown. Again, there is a clearly visible difference between the MOI results obtained here and in Ref. [8]. In all cases, the MOI spectra are very different from the exact spectra. This is especially apparent at the relatively small separation $\sigma/d = 0.05$ when the MOI predicts a spectral peak at $E \approx 1.5$Ev which is not present in the exact data. Even for the relatively large separation $\sigma/d = 0.35$, the MOI produces a two-peak structure, while the exact spectrum has only one peak. (We note here that in Fig. 3d of Ref. [8], the respective curve has also only one peak, but its maximum is about 10% smaller than the maximum of the spectrum in the noninteracting case. In the exact result, the maximum is approximately equal to that for the non-interacting case.)

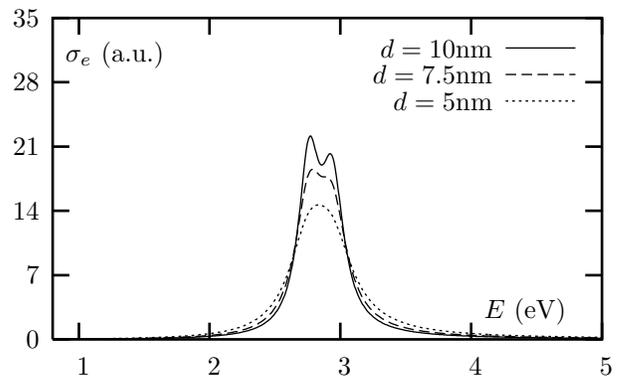

Fig. 4. The MOI result for the extinction spectrum of two spheres of different diameters $d$ and the ratio $\sigma/d = 0.35$. Polarization of the external field is parallel to the axis of symmetry.

The conclusion that can be made so far is that the MOI is inadequate for the spectral range and set of parameters used in Figs. 2,3 of Ref. [8]. The inaccuracy of the MOI is especially evident at smaller inter-sphere separations and for larger sphere diameters. We now discuss the possible cause of the discrepancy of the MOI calculations presented here and in Ref. [8]. In Fig. 4, we plot the MOI curve for $\sigma/d = 0.35$ and different values of $d$. The two-peak spectrum obtained at $d = 10$nm is the same as the one shown in Fig. 3d, while the single peak spectrum obtained at $d = 5$nm closely resembles the curve shown in Fig. 3d of Ref. [8]. Thus, the possible cause of the discrepancy is that in Ref. [8] the actual value of the sphere diameter used in calculations was twice smaller than what is shown in figure captions. That is, calculations in Fig. 2 of Ref. [8] were actually performed for $d = 2.5$nm and in Fig. 3 of Ref. [8] for $d = 5$nm. Under these circumstances, the relaxation due to the finite size effects is extremely strong and the spectral parameter $s$ has a large imaginary part which effectively weakens the multipole interaction of the spheres.

We now discuss the argument given in Ref. [10] that the MOI is accurate as long as the denominators in Eq. 5 do not vanish and the oscillator strengths $F_n$'s obey the appropriate sum rule. In general, this statement is correct when $|s + s_n| \gg 1$ for all $n$. But the result obtained from the MOI in this limit is, essentially, the non-interacting result. Therefore, application of the MOI under these conditions is simply not necessary. This is discussed in more detail below.

First, we note that $s$ is a complex variable while the depolarization factors $s_n$ are all real. The exact values of $s_n$ satisfy the inequality $0 < s_n < 1$, while the formulas (7),(8) can result in negative values of $s_n^{(L)}$ if $\sigma/d < (2^{1/3} - 1)^2 / 2^{4/3} \approx 0.0268$. It can be seen that the smallest possible value of $s_n^{(L)}$ that can be obtained from formula (7) is $-1/3$. If the complex value of $s$ is sufficiently separated from all $s_n$'s in the complex plane, one can replace the denominator $s - s_n$ by $s$. (In the meanfield approximation, the denominators are replaced by $s - Q$, where $Q$ is the appropriate average of the interaction operator [15].) The result is (taking into account the sum rule for $F_n$'s) the polarizability of an isolated (non-interacting) sphere. We can further expand the result in powers of the small parameter $s_n/s$ and thus obtain corrections to the non-interacting result. Unlike the former, these corrections depend on the particular choice of $F_n$ and $s_n$. An important point is that even if the corrections are small, they are not necessarily physically meaningful. I have previously demonstrated [9] that the corrections to the non-interacting polarizability obtained with the particular choice of $F_n$ and $s_n$ (6)-(8) are inaccurate for the ratio $\sigma/d$ smaller than $\approx 0.3$. Thus, any coincidences between the MOI and the exact spectra, such as the ones shown in Ref. [10], are due to the fact that the multipole interaction is very weak for the particular choice of parameters, and the obtained MOI spectra are, essentially, the spectra of non-interacting spheres. But in the spectral regions where interaction is essential, there is an obvious discrepancy between the MOI and the exact results [10].

We note that the complex spectral parameter $s$ can be removed from the section of the real axis occupied by the factors $s_n$, in particular, due to strong absorption. In that case, $s$ acquires a large imaginary part. This was the case for the simulations shown in Figs. 2,3 of Ref. [8]. Here the resonant interaction of the spheres was suppressed by finite size effects (2). In addition, the physical size of the spheres used in calculations appears to be smaller by the factor of 2 than what is claimed in figure captions. This resulted in relatively modest changes in the spectra (compared to the spectra of non-interacting spheres) which appear to be realistic. However, comparison with numerical results clearly demonstrates that even if the very strong relaxation is taken into account, the MOI results are qualitatively inaccurate.

Finally, it was suggested in Ref. [10] that at small electromagnetic frequencies (e.g., $f < 1 GHz$), the real part of the spectral parameter $s$ is always positive, so that the denominators $s + s_n$ can never vanish. This is clearly incorrect for both conductors and dielectrics. In the case of conductors, the real part of the dielectric function (1) is negative and of large magnitude. Then the real part of the spectral parameter $\epsilon_m/(\epsilon - \epsilon_m)$ is also negative and approaches zero from the left. It should be emphasized that in all realistic cases $\epsilon_h$ is of the order of unity and can not compensate for the term $-\omega_p^2/\omega^2$ in the spectral range $f < 1 GHz$. In the case of dielectrics, the low-frequency limit of the real part of $s$ can be either positive or negative, depending on the sign of $\epsilon - \epsilon_m$.